\documentclass{book}    
\usepackage{piers}  
\usepackage{amsmath,epsfig}
\usepackage{balance}
\usepackage{cite}
\usepackage{graphicx}

\usepackage{url}

\graphicspath{{Figures/}}
\usepackage{xcolor}
\usepackage{url}
\usepackage{float}

\usepackage{subfigure}
\usepackage{color}
\usepackage{balance}
\usepackage{float}
\usepackage{multirow}
\usepackage{soul}
\usepackage{tabularx}

\begin{document}

\title{Subjective Quality Assessment of Ground-based Camera Images}
\maketitle

\author      {F. M. Lastname}
\affiliation {University}
\address     {}
\city        {Boston}
\postalcode  {}
\country     {USA}
\phone       {345566}    
\fax         {233445}    
\email       {email@email.com}  
\misc        { }  
\nomakeauthor

\author      {F. M. Lastname}
\affiliation {University}
\address     {}
\city        {Boston}
\postalcode  {}
\country     {USA}
\phone       {345566}    
\fax         {233445}    
\email       {email@email.com}  
\misc        { }  
\nomakeauthor

\begin{authors}

{\bf Lucie L\'ev\^eque}$^{1}$, {\bf Soumyabrata Dev}$^{2,3,4*}$, {\bf Murhaf Hossari}$^{2}$, \\{\bf Yee Hui Lee}$^{5}$, {\bf and Stefan Winkler}$^{6}$\\
\medskip

$^{1}$Xi'an Jiaotong-Liverpool University, Suzhou, China\\

$^{2}$ADAPT SFI Research Centre, Dublin, Ireland\\

$^{3}$School of Computer Science, University College Dublin, Ireland.\\

$^{4}$UCD Earth Institute, University College Dublin, Ireland.\\

$^{5}$Nanyang Technological University Singapore, Singapore 639798\\

$^{6}$ Department of Computer Science, National University of Singapore, Singapore 117417\\

$^{*}$ Presenting author and corresponding author

\end{authors}


\begin{paper}

\begin{piersabstract}
Image quality assessment is critical to control and maintain the perceived quality of visual content. Both subjective and objective evaluations can be utilised, however, subjective image quality assessment is currently considered the most reliable approach. Databases containing distorted images and mean opinion scores are needed in the field of atmospheric research with a view to improve the current state-of-the-art methodologies. In this paper, we focus on using ground-based sky camera images to understand the atmospheric events. We present a new image quality assessment dataset containing original and distorted nighttime images of sky/cloud from SWINSEG database. Subjective quality assessment was carried out in controlled conditions, as recommended by the ITU. Statistical analyses of the subjective scores showed the impact of noise type and distortion level on the perceived quality.
\end{piersabstract}

\psection{Introduction}

With the tremendous advancement in the photogrammetric sensors, there has been a paradigm shift in the manner we analyse the atmosphere. Traditionally, satellite images were the only sources of visual information of the earth's atmosphere~\cite{arking1985retrieval}. However, these satellite images have low temporal and low spatial resolutions. Therefore, nowadays, researchers use ground-based imaging systems to understand the various events in the atmosphere. These imaging systems are popularly called Whole Sky Imagers (WSIs)~\cite{shields2013day}. The WSIs capture the images of the atmosphere at regular time intervals, and provide rich information about the clouds and other related atmospheric events.

Such ground-based sky cameras are generally equipped with a wide-angle lens, in order to capture the entire horizon. In addition to the inherent distortion due to the lens, several other noises get inadvertently incorporated in the captured sky/cloud images. This creates a plethora of problems for the remote sensing analysts, while processing the captured images during the post-analysis stage. During the curation and creation of sky/cloud image datasets, the remote sensing analysts choose the images carefully that are considered noise-free. However, no consideration is applied in understanding the impact of noise in the subjective quality of the considered sky/cloud images in the dataset. Such subjective assessment of a noisy sky/cloud image are generally ignored in the literature. Generally, the images in the popular sky/cloud image datasets are chosen carefully, such that they are noise-free. Dandini \emph{et al.} in \cite{dandini2019halo} used ground-based sky cameras to compute the scattering phase function (SPF) in cirrus clouds. In \cite{pawar2019detecting}, Pawar \emph{et al.} proposed a methodology to detect clear sky images in ground-based sky cameras. Also, Dev \emph{et al.} in \cite{dev2016ground} provided an overview of the various machine learning techniques in ground-based sky imaging. However, none of these works provide an analysis on the impact of noise in the subjective quality of the ground-based camera images. The visual distortions in the captured sky/cloud image can impact several post-analysis of sky/cloud images \emph{viz.} cloud coverage computation~\cite{7471439}, cloud-type recognition~\cite{7350833} and cloud-base height computation~\cite{8296259}. None of the existing works in the literature provide an understanding on the subjective image quality evaluation of the ground-based observation. And since the quality of the sky images have a direct impact on the information extracted, and also on decisions made based on the analysis of these images, we wanted to introduce a subjective assessment in this field. In this paper, we attempt to bridge this gap, by providing a subjective assessment of the quality of ground-based sky camera images. 

The use of image quality assessment is necessary to evaluate the quality of visual content~\cite{leveque2017study}. Two different approaches can be considered for image and video quality assessment, \emph{i.e.}, objective and subjective evaluations~\cite{MohammadiES14}. Objective assessment is based on mathematical algorithms which provide global or local quality measures. This method is reproducible and does not need human observation, as the main goal of an image or video quality assessment metric is to automatically predict the quality of a content. On the contrary, subjective image quality assessment requires observers for a visual study of perceived quality~\cite{LEVEQUE}~\cite{Aflaki2015}. As human observers are the ultimate receivers of visual content, subjective quality assessment is considered the most reliable approach. Furthermore, subjective data is needed to develop more accurate objective measures. The International Telecommunication Union (ITU) established standardised methods for subjective quality evaluation of images and videos~\cite{ITU}. In this paper, we present a new image quality database composed of $70$ stimuli, which consists of different types of noise and levels of degradation. Thirteen observers rated the quality of all images in controlled laboratory conditions.

The main contributions of this paper are as follows:
\begin{itemize}
    \item We provide the first-of-its-kind subjective image quality assessment of sky/cloud images captured by sky cameras;
    \item We propose and release a new image quality dataset comprising $70$ images with different noise levels; and 
    \item In the spirit of reproducible research, the source code used in generating the stimuli is available online\footnote{\url{https://github.com/Soumyabrata/cloud-quality}}.
\end{itemize}

\psection{Visual Quality Assessment}

\psubsection{Stimuli}
Figure~\ref{fig:illustration} illustrates the stimuli used for our experiment. It consists of ten nighttime images of sky scene. These ten images were randomly selected from SWINSEG (Singapore Whole sky Nighttime Imaging SEGmentation database)~\cite{dev2017nighttime} dataset. The images were captured using a ground-based whole sky imager, called Wide Angle High Resolution Sky Imaging System (WAHRSIS)~\cite{dev2014wahrsis}. This ground-based camera captures the images of the atmosphere at regular time intervals, with an interval of five minutes. It captures the images of the sky horizon throughout the $24$ hours of the day. In this paper, we focus our attention on \emph{only} nighttime sky/cloud images. Indeed, nighttime images are noisier in nature because of the high exposure time of the camera, and the low lighting conditions in the surroundings.

\begin{figure}[htb]
  \begin{center}
    \includegraphics[width=1\textwidth]{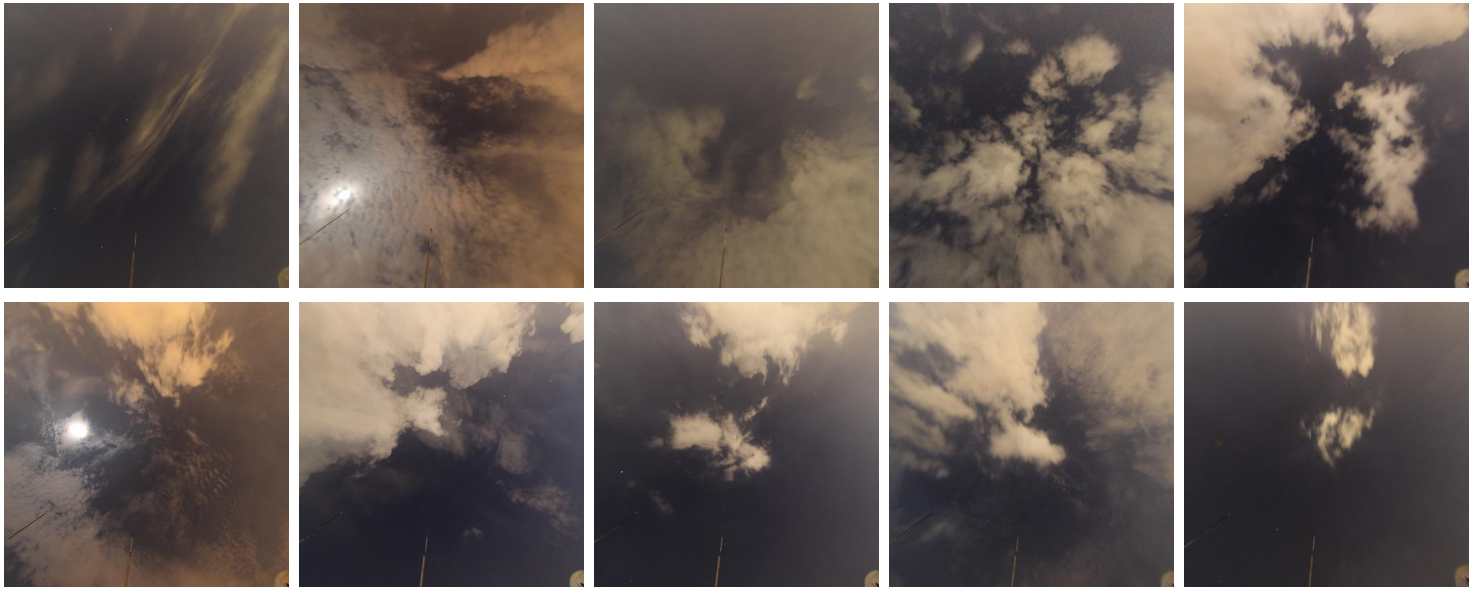}
  \end{center}
  \caption{Illustration of the original stimuli used in our experiment.}
  \label{fig:illustration}
\end{figure}

The ten original (\emph{i.e.}, reference) stimuli were distorted using three different types of image noise, \emph{i.e.}, Gaussian noise, salt-and-pepper noise, and speckle noise, at two different levels of distortion. We term the two different levels of noise as \emph{low} noise and \emph{high} noise, depending on the amount of added noise. We create noisy versions of the original stimuli by adding random noise to image with varying magnitude. In the case of images added with Gaussian noise, we add noise with varying degree of standard deviation ($\psi$) in the Gaussian distribution. We set $\psi = 5$ and $\psi = 12$ for low- and high- noisy images respectively. We also create noisy images that are noised with salt and pepper noise. In such noisy images, a random proportion of the pixels in a sky/cloud image are assigned to $0$ or $255$. The salt and pepper noisy images are generated only for grayscale images. We assign varying degree of probabilities to create the noisy version. We set this value of probability $p$, as $p=0.01$ and $p=0.03$ for low- and high- noisy images respectively. Finally, we also generate images noised with speckle noise. We define speckled noisy image as $\mathbf{\hat{X}} = \mathbf{X} + \mathbf{\Phi}.\mathbf{X}$, where $\mathbf{\hat{X}}$ is the noisy image, $\mathbf{X}$ is the reference image, and $\mathbf{\Phi}$ is a matrix of random Gaussian-distributed values with the same dimensions of the reference image. We also set varying values of $\sigma$ to generate the matrix $\mathbf{\Phi}$. We set $\psi = 0.25$ and $\psi = 0.5$ for the cases of low- and high- noisy images respectively.

\begin{figure}[htb]
\begin{center}
\makebox[0.75\textwidth][c]{\large {Images with Gaussian noise}}\\
\includegraphics[width=0.25\textwidth]{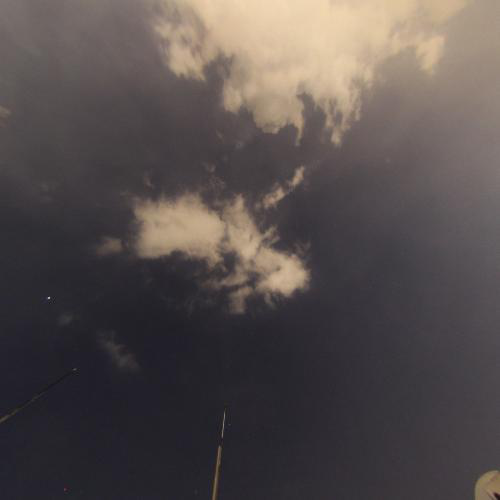}
\includegraphics[width=0.25\textwidth]{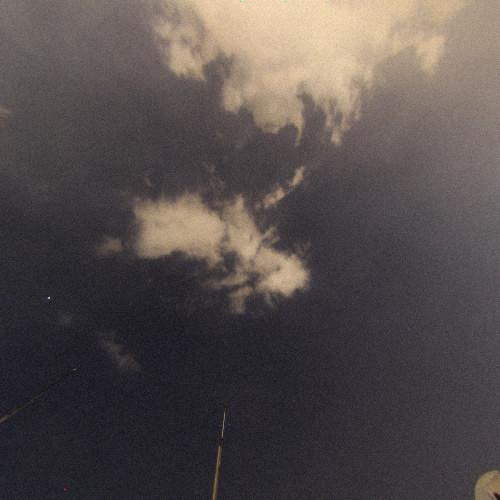}
\includegraphics[width=0.25\textwidth]{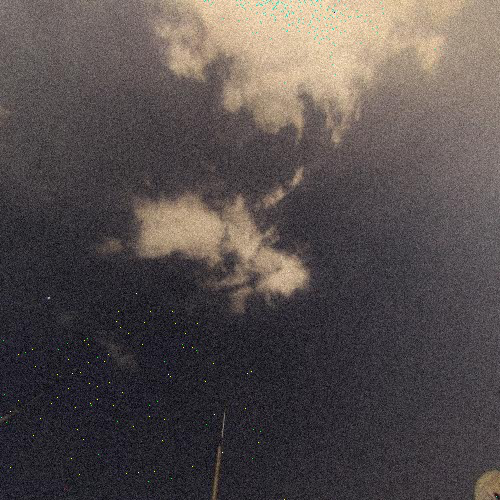}\\
\makebox[0.25\textwidth][c]{(a) Original image}
\makebox[0.25\textwidth][c]{(b) \texttt{+} low Gaussian noise.}
\makebox[0.25\textwidth][c]{(c) \texttt{+} high Gaussian noise.}\\
\hrulefill\\
\makebox[0.75\textwidth][c]{\large {Images with salt and pepper (S\&P) noise}}\\
\includegraphics[width=0.25\textwidth]{Fig2a.png}
\includegraphics[width=0.25\textwidth]{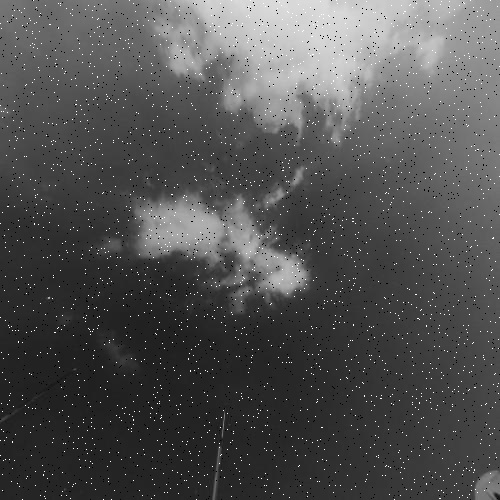}
\includegraphics[width=0.25\textwidth]{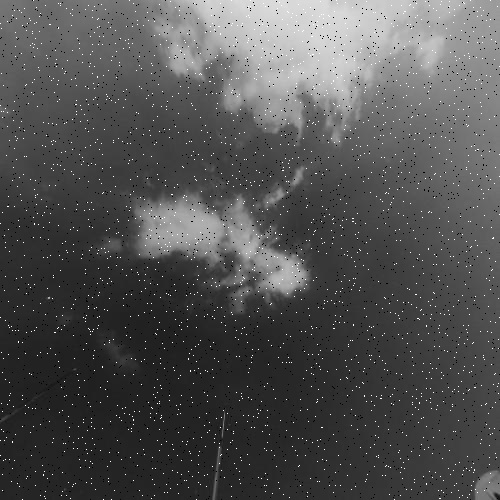}\\
\makebox[0.25\textwidth][c]{(d) Original image}
\makebox[0.25\textwidth][c]{(e) \texttt{+} low S\&P noise.}
\makebox[0.25\textwidth][c]{(f) \texttt{+} high S\&P noise.}\\
\hrulefill\\
\makebox[0.75\textwidth][c]{\large {Images with speckle noise}}\\
\includegraphics[width=0.25\textwidth]{Fig2a.png}
\includegraphics[width=0.25\textwidth]{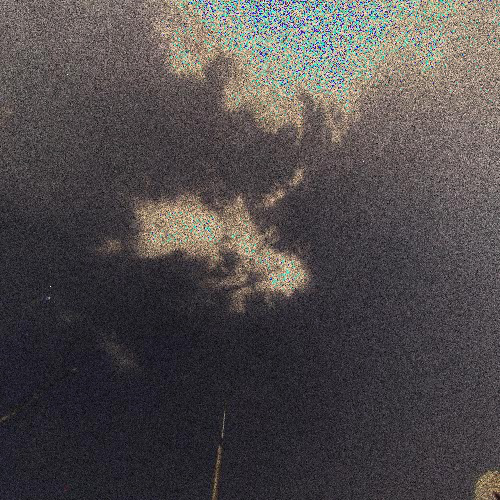}
\includegraphics[width=0.25\textwidth]{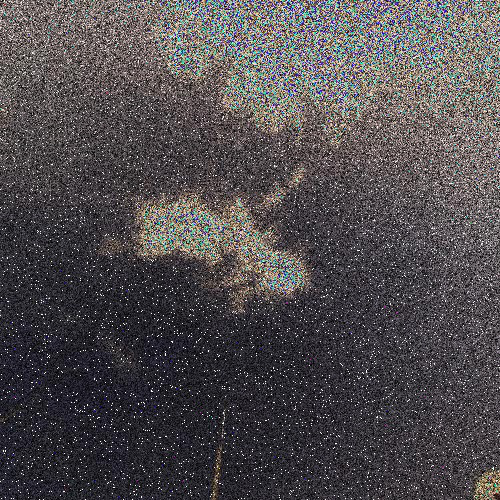}\\
\makebox[0.25\textwidth][c]{(g) Original image}
\makebox[0.25\textwidth][c]{(h) \texttt{+} low speckle noise.}
\makebox[0.25\textwidth][c]{(i) \texttt{+} high speckle noise.}
\caption{Illustration of various distortion configurations for a given nighttime sky/cloud image.
\label{fig:diff-levels}}
\end{center}
\end{figure}

Figure~\ref{fig:diff-levels} represents all the distortion configurations for a given sample stimulus. The salt and pepper noised images are defined only for grayscale images, whereas the Gaussian-noised and speckled noisy images are defined for RGB images. We can visually observe and discern the difference in the amount of added noise for both low- and high- noisy images.

\psubsection{Experimental Procedure}
We conducted a visual perception experiment using a multi-stimulus method~\cite{SAMVIQ}, where human subjects were asked to score the overall quality of the $70$ stimuli by inserting a slider mark on a continuous scale (i.e., from $0$ to $100$). The quality scale was divided into five semantic portions for scoring image quality: Bad ($0$-$20$), Poor ($20$-$40$), Fair ($40$-$60$), Good ($60$-$80$), and Excellent ($80$-$100$). Figure~\ref{fig:interface} illustrates the scoring interface used for the experiment. Figure~\ref{fig:interface} also represents an example of the test organisation for each source sequence where an explicit reference and six compressed versions placed in a different random order are included. Subjects were allowed to view and grade any stimulus in any order per scene; and each stimulus can be viewed and assessed as many times as the subject wishes (note the final score remained recorded).

\begin{figure}[htb]
  \begin{center}
    \includegraphics[width=0.85\textwidth]{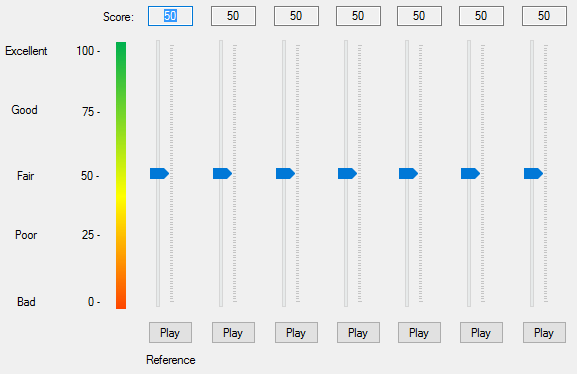}
  \end{center}
  \caption{Illustration of the interface used in our experiment.}
  \label{fig:interface}
\end{figure}

The experiment was conducted in a standard office environment at Xi'an Jiaotong-Liverpool University, Suzhou, China. The venue represented a controlled viewing environment to ensure consistent experimental conditions, \emph{i.e.}, low surface reflectance and approximately constant ambient light. The test stimuli were displayed on a $19$-inch LCD monitor screen, with a native resolution of $1920\times1080$ pixels. No image adjustment, \emph{e.g.}, zoom or window level, was allowed during the experiment.

Thirteen participants, including six men and seven women, participated in the experiment. Before the start of the actual experiment, each participant was provided with instructions on the procedure of the experiment (\emph{e.g.}, explaining the type of assessment and the scoring interface). A training session was conducted in order to familiarise the participants with the visual distortions involved and with how to use the range of the scoring scale. After training, all test stimuli were shown to each participant.

\psection{Experimental Results}
\psubsection{Data Pre-Processing}
To process the raw data, an outlier detection and subject exclusion procedure was applied to the scores~\cite{Sheikh2006}. An individual score would be considered an outlier if it was outside an interval of two standard deviations around the mean score for that image. A subject would be rejected if more than $20$\% of their scores were outliers. As a result of the above procedure, one participant was rejected from the database.

\psubsection{Impact of Distortion on Perceived Quality}

Figure~\ref{fig:res} illustrates the mean opinion score (MOS) averaged over all participants for each image in our experiment. It can be seen clearly that both the distortion type and distortion level seem to affect the perceived quality.

\begin{figure}[htb]
  \begin{center}
    \includegraphics[width=0.85\textwidth]{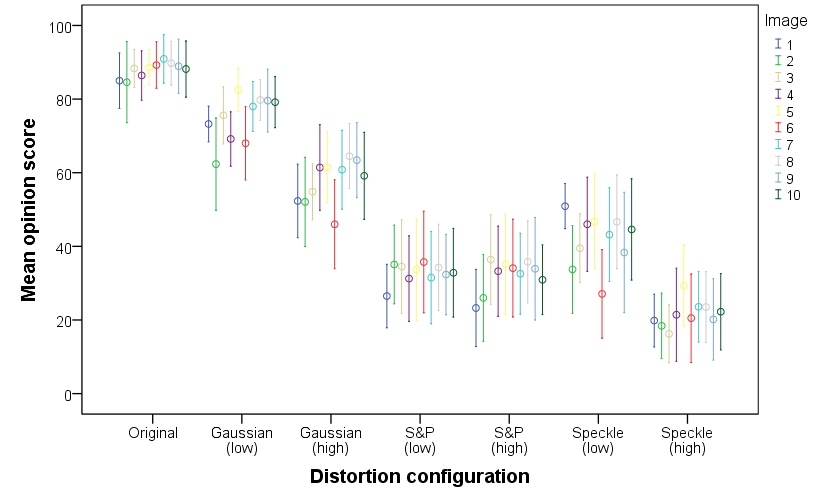}
  \end{center}
  \caption{Illustration of the mean opinion score (MOS) averaged over all subjects for each distorted image. Error bars indicate a 95\% confidence interval.}
  \label{fig:res}
\end{figure}

The observed tendencies are further statistically analysed using the Analysis of Variance (ANOVA). The results are summarised in Table~\ref{tab:table-name} where the F-statistic and its associated degrees of freedom and significance are included, along with the two-way interactions. The perceived quality is selected as the dependent variable, the image, distortion type, and distortion level as fixed independent variables, and the participant as random independent variable. The results show that there is no statistically significant difference between the participants in scoring the image quality (\emph{i.e.}, p$>0.05$). Due to the consistency in scoring, there is little need to calibrate the scores using z-scores as conventionally required in image quality~\cite{Hemminger1995}.

\begin{table}[htb]
\normalsize 
\centering 
\begin{center}
 \begin{tabular}{||c c c c||} 
 \hline
 Factor & df & F & p-value \\ [0.5ex] 
 \hline\hline
 Participant & 11 & 1.74 & 0.125 \\ 
 \hline
 Image & 9 & 1.96 & 0.110 \\
 \hline
 Distortion type & 2 & 9.89 & 0.029 \\
 \hline
 Distortion level & 1 & 4.39 & 0.174 \\
 \hline
 Distortion type * level & 2 & 23.38 & 0.000 \\ [1ex] 
 \hline
\end{tabular}
\newline
\newline
\caption
{\label{tab:table-name}Results of the ANOVA to evaluate the effects of `Participant', `Image', `Distortion type', and `Distortion level' on the perceived quality.}
\end{center}
\end{table}

\begin{figure}[htb]
  \begin{center}
    \includegraphics[width=0.85\textwidth]{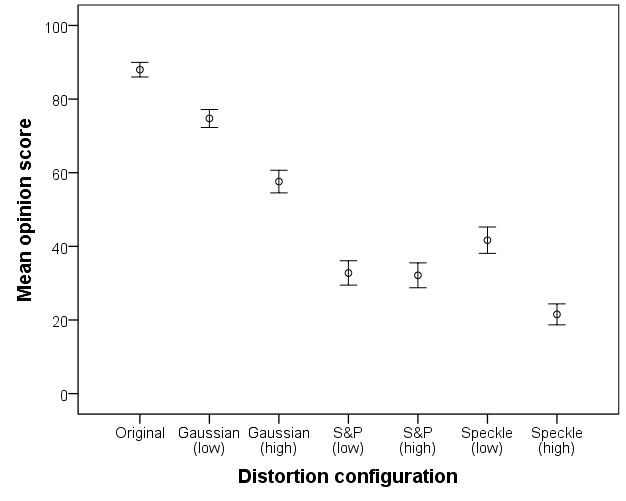}
  \end{center}
  \caption{Illustration of the mean opinion score (MOS) averaged over all subjects and all images. Error bars indicate a $95$\% confidence interval.}
  \label{fig:res2}
\end{figure}

 Similarly, the results show no statistically significant difference between the images (\emph{i.e.}, p$>0.05$). Figure~\ref{fig:res2} represents the mean opinion score averaged over all participants and over all images. The results of the ANOVA show a significant impact of the distortion configuration (\emph{i.e.}, distortion type and level) on the perceived quality (\emph{i.e.}, p$<0.005$). The post-hoc test reveals the following order in quality (note that commonly underlined entries are not significantly different from each other):
 
 Original ($<MOS> = 88.0$) $>$ Gaussian low ($<MOS> = 74.7$) $>$ Gaussian high ($<MOS> = 57.6$) $>$ Speckle low ($<MOS> = 41.7$) $>$ \ul{S\&P low ($<MOS> = 32.8$) $>$ S\&P high ($<MOS> = 32.1$)} $>$ Speckle high ($<MOS> = 21.5$).

It is interesting to notice that there is a statistically significant difference between the low and high distortion levels for both Gaussian noise and speckle noise, whereas there is no difference in scoring for the salt-and-pepper noise.

\psection{Conclusions and Discussion}
In this paper, we presented a new subjective image quality assessment database with $70$ nighttime/cloud images. Statistical analyses showed the impact of the distortion type, \emph{i.e.}, Gaussian noise, salt-and-pepper noise, and speckle noise on the human perceived quality, as well as the impact of the distortion level as far as Gaussian noise and speckle noise are concerned.

This work can be further used as a better understanding of the impact of visual distortion on ground-based camera images and, subsequently, improve the current sky imaging detection methodologies. We will also evaluate the impact of noise in the computation of cloud coverage and cloud-type recognition frameworks. 

\ack
The ADAPT Centre  for  Digital  Content  Technology  is  funded  under  the  SFI Research Centres Programme (Grant 13/RC/2106) and is co-funded under the European Regional Development Fund.

\bibliographystyle{IEEEbib}

\begin{thebibliography}{10}

\bibitem{arking1985retrieval}
A.~Arking and J.~D. Childs,
\newblock ``Retrieval of cloud cover parameters from multispectral satellite
  images,''
\newblock {\em Journal of Climate and Applied Meteorology}, vol. 24, no. 4, pp.
  322--333, 1985.

\bibitem{shields2013day}
J.~E. Shields, M.~E. Karr, R.~W. Johnson, and A.~R. Burden,
\newblock ``Day/night whole sky imagers for 24-h cloud and sky assessment:
  history and overview,''
\newblock {\em Applied optics}, vol. 52, no. 8, pp. 1605--1616, 2013.

\bibitem{dandini2019halo}
P.~Dandini, Z.~Ulanowski, D.~Campbell, and R.~Kaye,
\newblock ``Halo ratio from ground-based all-sky imaging,''
\newblock {\em Atmospheric Measurement Techniques}, vol. 12, no. 2, pp.
  1295--1309, 2019.

\bibitem{pawar2019detecting}
P.~Pawar, C.~Cort\'{e}s, K.~Murray, and J.~Kleissl,
\newblock ``Detecting clear sky images,''
\newblock {\em Solar Energy}, vol. 183, pp. 50 -- 56, 2019.

\bibitem{dev2016ground}
S.~Dev, B.~Wen, Y.~H. Lee, and S.~Winkler,
\newblock ``Ground-based image analysis: A tutorial on machine-learning
  techniques and applications,''
\newblock {\em IEEE Geoscience and Remote Sensing Magazine}, vol. 4, no. 2, pp.
  79--93, 2016.

\bibitem{7471439}
S.~{Dev}, Y.~H. {Lee}, and S.~{Winkler},
\newblock ``Color-based segmentation of sky/cloud images from ground-based
  cameras,''
\newblock {\em IEEE Journal of Selected Topics in Applied Earth Observations
  and Remote Sensing}, vol. 10, no. 1, pp. 231--242, Jan 2017.

\bibitem{7350833}
S.~{Dev}, Y.~H. {Lee}, and S.~{Winkler},
\newblock ``Categorization of cloud image patches using an improved
  texton-based approach,''
\newblock in {\em Proc. IEEE International Conference on Image Processing
  (ICIP)}, Sep. 2015, pp. 422--426.

\bibitem{8296259}
F.~M. {Savoy}, S.~{Dev}, Y.~H. {Lee}, and S.~{Winkler},
\newblock ``Stereoscopic cloud base reconstruction using high-resolution whole
  sky imagers,''
\newblock in {\em 2017 IEEE International Conference on Image Processing
  (ICIP)}, Sep. 2017, pp. 141--145.

\bibitem{leveque2017study}
L.~L{\'e}v{\^e}que, W.~Zhang, C.~Cavaro-M{\'e}nard, P.~Le~Callet, and H.~Liu,
\newblock ``Study of video quality assessment for telesurgery,''
\newblock {\em IEEE Access}, vol. 5, pp. 9990--9999, 2017.

\bibitem{MohammadiES14}
P.~Mohammadi, A.~Ebrahimi{-}Moghadam, and S.~Shirani,
\newblock ``Subjective and objective quality assessment of image: {A} survey,''
\newblock {\em CoRR}, vol. abs/1406.7799, 2014.

\bibitem{LEVEQUE}
L~L{\'e}v{\^e}que, H~Liu, S~Barakovi{\'c}, J~Barakovi{\'c}~Husi{\'c},
  M~Martini, M~Outtas, L~Zhang, A~Kumcu, L~Platisa, R~Rodrigues, A~Pinheiro,
  and A~Skodras,
\newblock ``On the subjective assessment of the perceived quality of medical
  images and videos,''
\newblock {\em Tenth International Conference on Quality of Multimedia
  Experience (QoMEX)}, 05 2018.

\bibitem{Aflaki2015}
P.~Aflaki, M.~M. Hannuksela, and M.~Gabbouj,
\newblock ``Subjective quality assessment of asymmetric stereoscopic {3D}
  video,''
\newblock {\em Signal, Image and Video Processing}, vol. 9, no. 2, pp.
  331--345, Feb 2015.

\bibitem{ITU}
International Telecommunication~Union (ITU),
\newblock ``Subjective video quality assessment methods for multimedia
  applications,''
\newblock {\em Recommendation ITU-T P.910}, 04 2008.

\bibitem{dev2017nighttime}
S.~Dev, F.~M. Savoy, Y.~H. Lee, and S.~Winkler,
\newblock ``Nighttime sky/cloud image segmentation,''
\newblock in {\em Proc. IEEE International Conference on Image Processing
  (ICIP)}. IEEE, 2017, pp. 345--349.

\bibitem{dev2014wahrsis}
S.~Dev, F.~M. Savoy, Y.~H. Lee, and S.~Winkler,
\newblock ``{WAHRSIS}: A low-cost high-resolution whole sky imager with
  near-infrared capabilities,''
\newblock in {\em Infrared Imaging Systems: Design, Analysis, Modeling, and
  Testing XXV}. International Society for Optics and Photonics, 2014, vol.
  9071, p. 90711L.

\bibitem{SAMVIQ}
F.~Kozamernik, V.~Steinmann, P.~Sunna, and E.~Wyckens,
\newblock ``{SAMVIQ} - a new {EBU} methodology for video quality evaluations in
  multimedia,''
\newblock {\em SMPTE Motion Imaging Journal}, vol. 114, pp. 152--160, 04 2005.

\bibitem{Sheikh2006}
H.~R. Sheikh, M.~F. Sabir, and A.~C. Bovik,
\newblock ``A statistical evaluation of recent full reference image quality
  assessment algorithms,''
\newblock {\em Trans. Img. Proc.}, vol. 15, no. 11, pp. 3440--3451, Nov. 2006.

\bibitem{Hemminger1995}
Bradley~M. Hemminger, R.~Eugene Johnston, Jannick~P. Rolland, and Keith~E.
  Muller,
\newblock ``Introduction to perceptual linearization of video display systems
  for medical image presentation,''
\newblock {\em Journal of Digital Imaging}, vol. 8, no. 1, pp. 21--34, Feb
  1995.

\end{thebibliography}

\end{paper}

\end{document}